\documentclass[12pt]{article}
\pdfoutput=1

\usepackage{latexsym}
\usepackage{epsfig,amssymb,euscript}
\usepackage{amsmath}
\usepackage{mathrsfs}
\usepackage{ccaption}             
\usepackage[usenames]{color}

\usepackage[
      colorlinks=true,
      linkcolor=blue,
      urlcolor=blue,
      filecolor=black,
      citecolor=red,
      pdfstartview=FitV,
      pdftitle={},
        pdfauthor={Benjamin Withers},
        pdfsubject={},
        pdfkeywords={},
        pdfpagemode=None,
        bookmarksopen=true
      ]{hyperref}

\def\be{\begin{equation}}
\def\ee{\end{equation}}
\def\bea{\begin{eqnarray}}
\def\eea{\end{eqnarray}}

\begin{document}

\makeatletter
\renewcommand{\theequation}{\thesection.\arabic{equation}}
\@addtoreset{equation}{section}
\makeatother

\baselineskip 18pt

\def\todo#1{\textbf{[}\emph{\textbf{#1}}\textbf{]}}

\begin{titlepage}

\vfill

\begin{flushright}
\end{flushright}

\vfill

\begin{center}
   \baselineskip=16pt
   {\Large\bf Black branes dual to striped phases}
   %
  \vskip 1.5cm
      Benjamin Withers\\
   \vskip .6cm
     \textit{Centre for Particle Theory and Department of Mathematical Sciences\\ Durham University, South Road, Durham, DH1 3LE, U.K.}


\end{center}

\vfill

\begin{center}
\textbf{Abstract}
\end{center}

\begin{quote}
We construct inhomogeneous charged black branes in AdS, holographically dual to a phase at finite chemical potential with spontaneously broken translation invariance in one direction. These are obtained numerically, solving PDEs for the fully backreacted system. Fixing the periodicity scale, we find a second order phase transition to the inhomogeneous phase. We comment on the properties of the state emerging at low temperatures. For some models we demonstrate the existence of a branch of striped solutions but no continuous phase transition.
\end{quote}

\vfill

\end{titlepage}

\section{Introduction}
An approach to applied holography is to pick a gravitational bulk theory, either in a consistent truncation of supergravity or a phenomenological model, and search for solutions which describe various equilibrium phases of the dual field theory. Guided by condensed matter phenomena this strategy has proven to be a fruitful approach and has driven the construction of  a variety of novel black brane solutions. 

One well studied class of solutions are those dual to holographic superfluids. In the simplest case a charged scalar field is added to the bulk model resulting in a branch of black branes with scalar hair, emerging from the charged Reissner-Nordstrom (RN) solution~\cite{Gubser:2008px,Hartnoll:2008vx,Hartnoll:2008kx}. If this branch is thermodynamically preferred then the dual field theory will spontaneously break its global U(1) symmetry. The nature of the phase transition to the broken phase, if indeed there is one at all, depends strongly on the details of the bulk model away from the threshold temperature~\cite{Donos:2011ut}. Such phases are not restricted to phenomenological models, existing in consistent truncations of supergravity Kaluza-Klein reductions~\cite{Gauntlett:2009dn,Gubser:2009qm}.

Another class of phases are those in which translational invariance of an electrically charged phase is spontaneously broken \cite{Nakamura:2009tf,Ooguri:2010kt,Donos:2011bh,Donos:2011ff}.  Such phases are also seen for magnetically charged black branes \cite{Donos:2011qt,Donos:2011pn} and occur for a wide range of bulk models, recent examples include massive gravity \cite{Vegh:2013sk} and Einstein-Maxwell-dilaton systems \cite{Donos:2013gda}.\footnote{See also \cite{Domokos:2007kt,Ooguri:2010xs,Bayona:2011ab,Bergman:2011rf} for spatial modulation in a probe brane setting.} In other examples, the instability can result in homogeneous but non-translationally invariant stationary black brane solutions, which can be constructed by solving ODEs \cite{Iizuka:2012iv,Donos:2012gg,Donos:2012wi}. In the case of \cite{Donos:2012gg,Donos:2012wi}, the phase transition was found to be second order, leading to a modulated zero entropy state at zero temperature.

In this paper we will focus on models of the type analysed linearly in \cite{Donos:2011bh}, which break translations resulting in a non-homogeneous phase. The key ingredient in this model is a parity violating coupling with a pseudo-scalar field. A zero mode analysis indicates that this term is responsible for a new phase emanating from an electrically charged black brane, displaying both charge and current density waves. Moving beyond the linear analysis in these cases is desirable, principally to determine if there is a phase transition. Once a solution has been obtained one can begin to address issues such as the nature of the ground state, the number of solutions which can exist at a given temperature, the thermodynamically preferred spatial scale, transport properties, \emph{etc}.

In this spirit, we set about the construction of cohomogeneity-two charged black brane solutions with AdS$_4$ asymptotics. Whilst the linear analysis performed in \cite{Donos:2011bh} has the same cohomogeneity (hence `stripes'), it does so without loss of generality; the zero modes may be simply superposed in different directions leading to solutions with lower symmetry. Thus we should expect cohomogeneity-three branches of solution emerging from the threshold temperature for linear stability, and ultimately these may turn out to dominate the ensemble. Here, for technical simplicity, we choose to focus on the special case of a cohomogeneity-two family of solutions, translationally invariant in one boundary direction. This requires the solution of a 2d PDE boundary value problem involving the Einstein equations. Our approach will utilise elegant methods~\cite{Headrick:2009pv,Adam:2011dn,Wiseman:2011by} for rendering a problem such as this elliptic.  In a technically similar but physically distinct scenario these methods have been employed in the construction of a variety of holographic lattices forced by UV boundary conditions~\cite{Horowitz:2012ky,Horowitz:2012gs,Horowitz:2013jaa}.

We note that solutions of the type constructed here have previously been sought in the work~\cite{UBC} employing a different numerical method and finding a  first-order phase transition at temperatures lower than the threshold of linear instability.
\\

The organisation and key results of this paper are as follows. In sections \ref{setup} and \ref{ansatz} we specify the bulk model, ansatz and numerical method. In section \ref{repsec} we present a single striped solution. In section \ref{secondsec} we examine the temperature dependence at fixed periodicity, $2\pi\mu/k$, finding second order phase transitions and evidence of a zero entropy inhomogeneous state at zero temperature.
In section \ref{secfirst} we present an example of a different model, identical in linear theory, which does not exhibit a continuous phase transition. That is, as with the holographic superfluid constructions, the behaviour away from $T_c$ is strongly dependent on the details of the bulk model. We conclude in section \ref{discussion}.
\\\\
\emph{Note added:} While this paper was being finalised \cite{DonosSolo} appeared, also finding second order phase transitions.

\section{The bulk theory}\label{setup}
We consider the bulk action
\be
S_{\text{bulk}}= \int d^4x \sqrt{-g}\left(R -\frac{1}{2}\left(\partial \phi\right)^2 - \frac{\tau(\phi)}{4}F^2 - V(\phi)\right) - \int \frac{\vartheta(\phi)}{2}F\wedge F,\label{sbulk}
\ee
where $F=d\mathcal{A}$ and we have set $16\pi G=1$.\footnote{Note that to make contact with the linearised instabilities of \cite{Donos:2011bh}, we can identify $T/\mu_{\text{here}} = \frac{1}{2}T/\mu_{\text{there}}$ as well as $k/\mu_{\text{here}} = \frac{1}{\sqrt{2}}k/\mu_{\text{there}}$. We have kept an equivalent linearised parameterisation of $\tau(\phi),  V(\phi)$ and $\vartheta(\phi)$ such that $n,c_{1{here}} = n,c_{1{there}}$.} This action will be supplemented later by Gibbons-Hawking and boundary counter-terms.
We choose to work with functions $\tau(\phi),  V(\phi)$ and $\vartheta(\phi)$ which admit electrically charged RN black brane solutions with an asymptotic ($L=1$) AdS$_4$, when $\phi = 0$. With this condition we may conveniently parameterise the functions in the vicinity of the phase transition in a small $\phi$ expansion 
$\tau = \hat{\tau}+O(\phi)^4$,
$V=\hat{V}+O(\phi)^4$ and 
$\vartheta = \hat{\vartheta}+O(\phi)^3$ 
where  
$\hat{\tau} \equiv 1- \frac{n}{24}\phi^2$, 
$\hat{V}\equiv-6+\frac{1}{2}m^2\phi^2$ 
and $\hat{\vartheta} \equiv \frac{c_1}{2\sqrt{6}}\phi$. 
Note that such a description will only hold in the vicinity of the critical temperature and will not suffice for example, in determining the order of the phase transitions.

Throughout this paper we will consider only the case $m^2=-2$, where we take the operator dual to the scalar $\phi$, $\mathcal{O}_\phi$,  to have dimension-2, setting its dimension-1 source to zero. For concreteness we will consider two phenomenological models,
\bea
\tau &=& \text{sech}(\sqrt{3}\phi),\quad V = -6\,\text{cosh}(\phi/\sqrt{3}), \quad \vartheta = \frac{c_1}{6\sqrt{2}}\,\text{tanh}(\sqrt{3}\phi).\label{nlmodel}\\
\tau &=& \hat{\tau},\quad V = \hat{V}, \quad \vartheta = \hat{\vartheta}.\label{linmodel}
\eea
The specific case of \eqref{linmodel} with $n=0$ and $c_1 =4.5$ was considered in \cite{UBC}. The nonlinear model \eqref{nlmodel} when $c_1=6\sqrt{2}$ appears in a consistent truncation of a reduction on an arbitrary Sasaki-Einstein space, $SE_7$ \cite{Gauntlett:2009bh,Gauntlett:2009zw,Donos:2012yu}. Here we consider a phenomenological modification of this model by dialling the value of $c_1$ away from $6\sqrt{2}$. In practice we increase $c_1$  a small amount to $c_1=9.9$. This has the advantage of raising the critical temperature for improved numerical accuracy.
\section{Ansatz and implementation\label{ansatz}}
To construct the striped black branes we retain diffeomorphism invariance in the $x$ and $z$ directions, adopting the following stationary ansatz,
\bea
ds^2 &=& \frac{1}{z^2}\left(-T f(z)dt^2 + Z\frac{dz^2}{f(z)}+X (dx+\gamma dz)^2 + Y(dy+\beta dt)^2\right)\\
\mathcal{A} &=& A dt + B dy
\eea
where $T,Z,X,Y,\gamma,\beta, A, B$ and $\phi$ are functions both of the AdS radial coordinate, $z$, and a single spatial boundary direction $x$. The ansatz is translationally invariant in the remaining spatial boundary coordinate, $y$. 
%
The function $f(z) \equiv (1-z)(1+z+z^2-\mu^2z^3/4)$ conveniently factors out the  electrically charged RN solution at $Z=T=X=Y=1$ and $\gamma=\beta=B=\phi=0$ and 
$A= \mu(1-z)$. This RN solution will constitute the normal phase of our system, with  temperature $T = (12-\mu^2)/(16\pi)$ and entropy density $s = 4\pi$. We have chosen coordinates such that $z=1$ gives the position of the outer horizon. 

\subsection{A regular radial coordinate}
For presentational purposes the Schwarzschild-like $z$-coordinate is efficient, and we find that it is also sufficient for numerical purposes, however, by switching to a radial coordinate regular on the horizon we achieve improved numerical accuracy at low temperatures. Consequently in what follows we will adopt the radial coordinate $r$, defined by,
\be
(1-r)^2 \equiv 1-z.
\ee
The black brane horizon is at $r=z=1$ and the boundary at $r=z=0$. 

We seek solutions with a regular horizon. This requires that the functions $Z,T, X,Y$, $\beta,A,B$, and $\phi$ are smooth functions of $(1-r)^2$ there, whilst the $dx\,dr$ cross term, $\gamma$ leads with $(1-r)$, but such that $\gamma/(1-r)$ is a smooth function of $(1-r)^2$. In addition, we choose the Killing vector $\partial_t$ to be null on the horizon; accordingly both $A$ and $\beta$ are taken to vanish there.\footnote{We could have also allowed for (spatially modulated or homogeneous) boosted black brane solutions by setting $\beta$ to a non-vanishing constant on the horizon, together with a suitable modified regularity condition for $A$. This would correspond to a Killing horizon for a linear combination of the Killing vectors $\partial_t$ and $\partial_y$, and extend the family of solutions by one additional parameter.} These conditions are imposed by factoring out appropriate powers of $(1-r)$ and using Neumann boundary conditions, for example, $\partial_r B=0$ at $r=1$. Finally we have the additional Dirichlet condition $T=Z$ at the horizon, rendering the surface gravity constant, determined by $f(1)$. In particular the temperature of the inhomogeneous solutions will be the same as RN at the same $\mu$, \emph{i.e.} $T = (12-\mu^2)/(16\pi)$. 

At the boundary we fix the boundary metric to be Minkowski and turn off all sources. Specifically we impose the Dirichlet boundary conditions $Z=T=X=Y=1$, $\beta = \gamma = B = \varphi= 0$ where we have now defined $\phi \equiv r \varphi$ and $A = \mu$ at $r=0$.

These boundary conditions lead to a two parameter family of solutions characterised by the dimensionless temperature $T/\mu$ and periodicity $k/\mu$. In particular there is no other freely specifiable data and consequently any spatial modulation will be spontaneous. 

\subsection{Implementation}
We may formulate the equations of motion resulting from \eqref{sbulk} as an elliptic, boundary value problem by utilising the methods \cite{Headrick:2009pv,Adam:2011dn,Wiseman:2011by}. In short, the problem may be rendered elliptic by considering the Harmonic Einstein equations, obtained by taking the field equations of  \eqref{sbulk} and making the replacement,
\be
R_{ab} \to R^{\text{H}}_{ab} =R_{ab} - \nabla_{(a} \xi_{b)}
\ee
introducing the vector field $\xi^a \equiv g^{mn}(\Gamma^a_{mn} - \tilde{\Gamma}^a_{mn})$, where $\Gamma$ is the Levi-Civita connection of $g$ and $\tilde{\Gamma}$ is the Levi-Civita connection of a reference metric $\tilde{g}$. Here, for simplicity,  we take $\tilde{g}$ to be the metric of the RN solution at the same value of $\mu$. To ensure that we have a solution to the equations of motion of \eqref{sbulk} we must verify that the vector $\xi$ is zero to the required numerical accuracy.

The resulting nine coupled Harmonic Einstein, Maxwell and scalar field equations are solved numerically using a Newton-Raphson method. The fields and their derivatives are represented using a spectral collocation method.\footnote{See \cite{Tref} for a pedagogical exposition of these methods.} We employ Chebyshev polynomial interpolants on an irregular grid for the $r$ coordinate containing $N+1$ points in the interval $[0,1]$, and Fourier interpolants on a regular grid for the $x$ coordinate containing $N$ points in the interval $[-\pi/k,\pi/k)$ with $x=-\pi/k$ and $x=\pi/k$ identified. It will be convenient to denote quantities averaged over the period with a bar, \emph{i.e.}
\be
\bar{F} \equiv \frac{k}{2\pi} \int^{\pi/k}_{-\pi/k} F(x) dx.\label{average}
\ee

We verify that our solutions are not Ricci-Solitons (\emph{i.e.} that they are solutions to the equations of motion of  \eqref{sbulk}) by monitoring the field $|\xi^2|$, ensuring $|\xi^2|<10^{-10}$ is satisfied at all grid points for the solutions presented in this paper. In practice higher values of $N$ are required to achieve this condition at low temperatures, and so for the majority of solutions presented in this paper we achieve values which are smaller than this limit by several orders of magnitude. Tests of the numerics are presented in appendix \ref{numericaltests}, in particular, in section \ref{nconv} we verify that $|\xi^2|$ converges to zero exponentially with $N$.

As a further check, we have also constructed some of these solutions using fourth-order finite differencing, finding consistent results.

\section{A representative solution\label{repsec}}

In this section we present a single representative striped black brane solution, for the model \eqref{nlmodel} with $c_1=9.9$.
As an initial check of our numerics, we verify that the inhomogeneous phase meets the RN branch along a `bell curve' of critical temperatures with good agreement with the $n=36$, $c_1\simeq9.9$ linear instability case constructed in \cite{Donos:2011bh}.

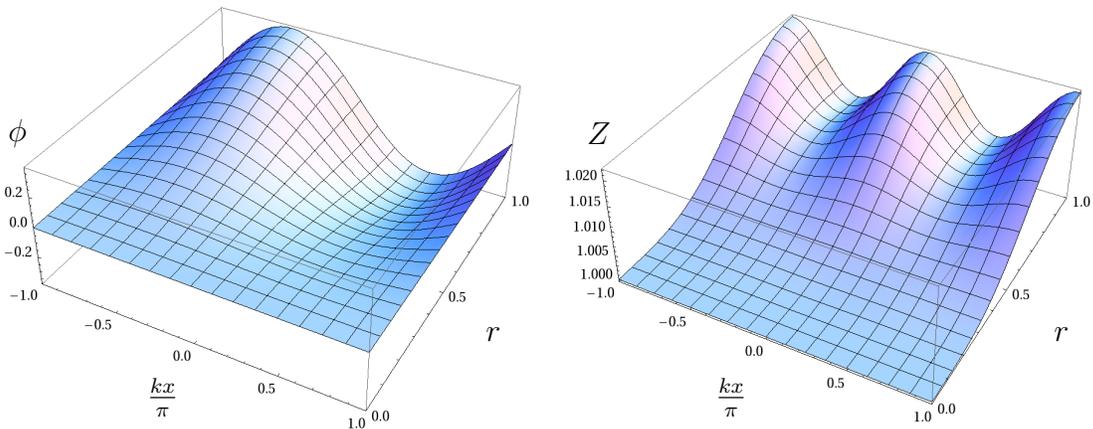
\begin{figure}[h!]
\begin{center}
\includegraphics[width=0.95\textwidth]{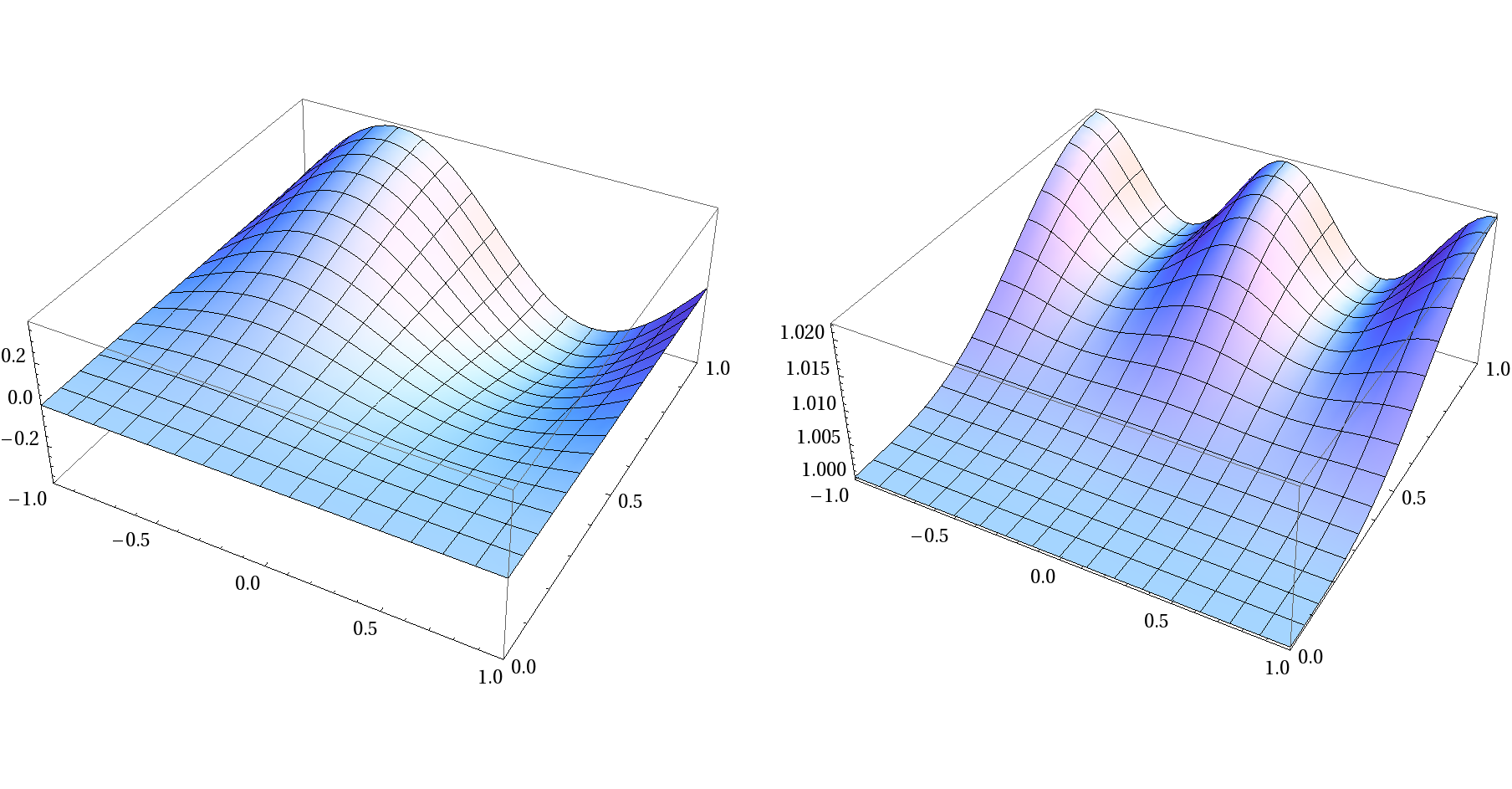}
\begin{picture}(0.1,0.1)(0,0)
\put(-355,40){\makebox(0,0){$\frac{kx}{\pi}$}}
\put(-140,40){\makebox(0,0){$\frac{kx}{\pi}$}}
\put(-230,64){\makebox(0,0){$r$}}
\put(-15,64){\makebox(0,0){$r$}}
\put(-410,140){\makebox(0,0){$\phi$}}
\put(-190,140){\makebox(0,0){$Z$}}
\end{picture}
\vskip 1em
\caption{Bulk profiles for the pseudo-scalar $\phi$ and a metric function, $Z$, for representative striped black brane solution in the model \eqref{nlmodel}, with $c_1=9.9$ and  $k/\mu=1.1/\sqrt{2}$ at $T/\mu \simeq 0.87 (T/\mu)_c$. The $r$-coordinate runs from $r=0$ on the boundary to $r=1$ at the horizon.
  \label{repsol}}
\end{center}
\end{figure}

We focus here on a single solution in the one-parameter family of solutions emerging from the linear bell-curve at $k/\mu=1.1/\sqrt{2}$, for which the critical temperature is given by $(T/\mu)_c \simeq 0.0236$.  We select a solution at a temperature of $T/\mu = 0.0205 \simeq 0.87 (T/\mu)_c$. It is thermodynamically preferred over  the RN solution. Bulk plots of the scalar $\phi$ and metric function $Z$ for this solution are presented in figure \ref{repsol}. 

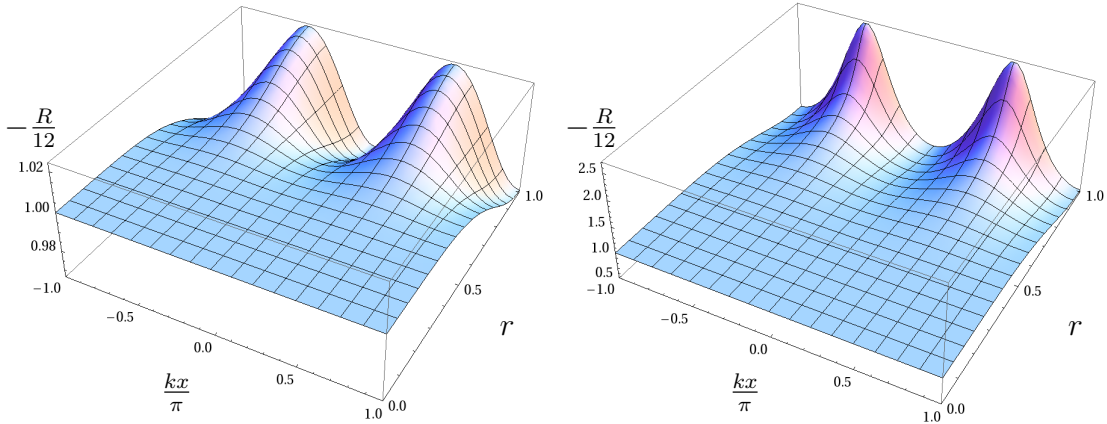
\begin{figure}[h!]
\begin{center}
\includegraphics[width=0.95\textwidth]{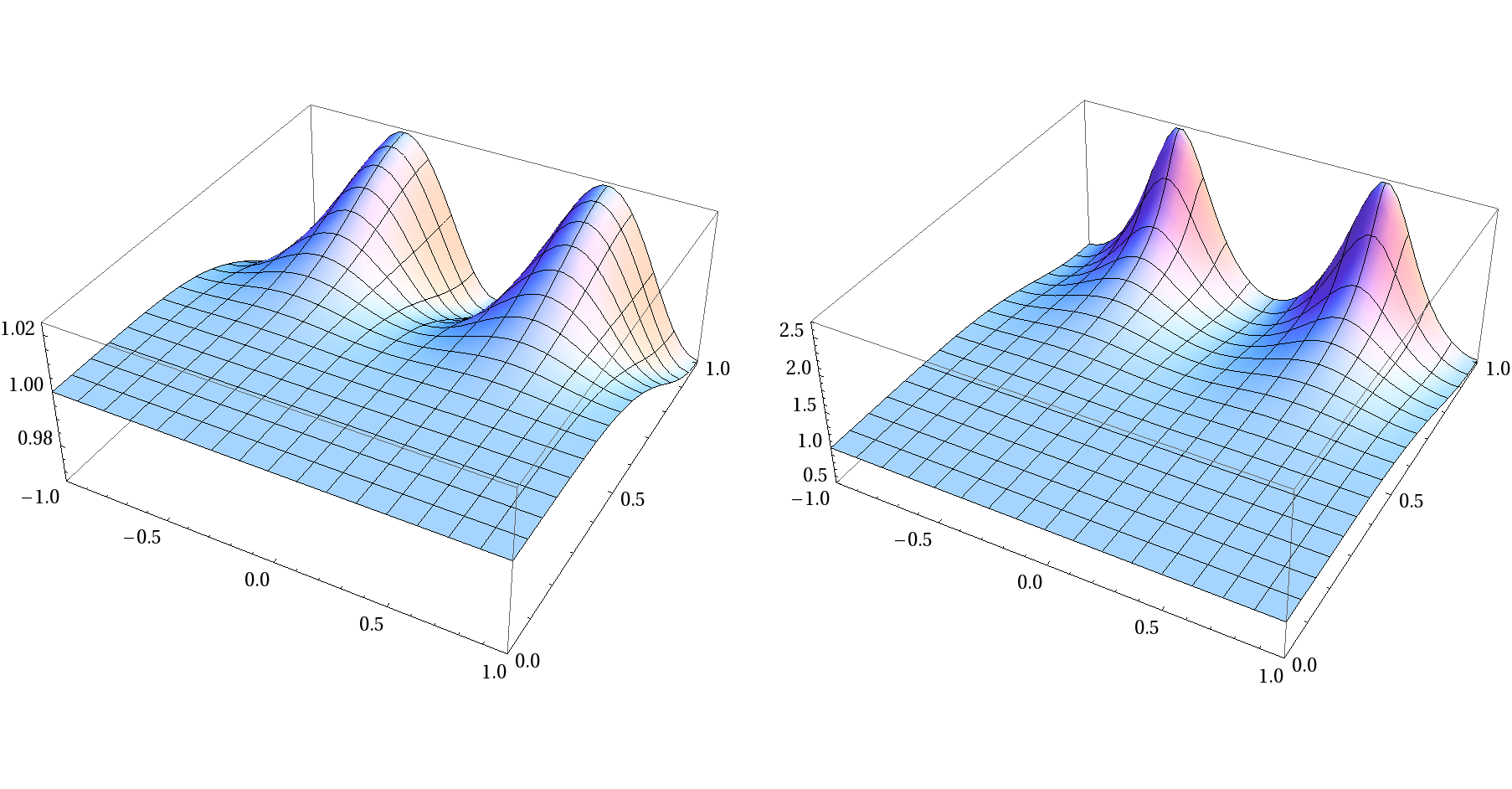}
\begin{picture}(0.1,0.1)(0,0)
\put(-355,40){\makebox(0,0){$\frac{kx}{\pi}$}}
\put(-140,40){\makebox(0,0){$\frac{kx}{\pi}$}}
\put(-230,64){\makebox(0,0){$r$}}
\put(-15,64){\makebox(0,0){$r$}}
\put(-410,140){\makebox(0,0){$-\frac{R}{12}$}}
\put(-198,140){\makebox(0,0){$-\frac{R}{12}$}}
\end{picture}
\vskip 1em
\caption{Ricci scalar in the bulk. \emph{Left:} The solution shown in figure \ref{repsol}, at at $T/\mu \simeq 0.87 (T/\mu)_c$. \emph{Right:} The same model and branch but at $T/\mu \simeq 0.04 (T/\mu)_c$.
  \label{ricci}}
\end{center}
\end{figure}

For this solution the Ricci scalar is shown in the left panel of figure \ref{ricci}, changing from $R=-12$ in the asymptotic AdS$_4$ and becoming spatially modulated on the horizon. For comparison in the right panel of figure \ref{ricci} we show the Ricci scalar for a solution on the same branch at much lower temperatures, $T/T_c\simeq 0.04$, exhibiting a much larger variation on the horizon.

\section{A branch at fixed $k/\mu$\label{secondsec}}
In this section we fix the periodicity of the solution, $k/\mu$ and examine the free energy as a function of temperature. As in section \ref{repsec} we employ the model \eqref{nlmodel}, with $c_1=9.9$ and we focus on the one-parameter family at $k/\mu=1.1/\sqrt{2}$.

\begin{figure}[h!]
\begin{center}
\includegraphics[width=0.4\textwidth]{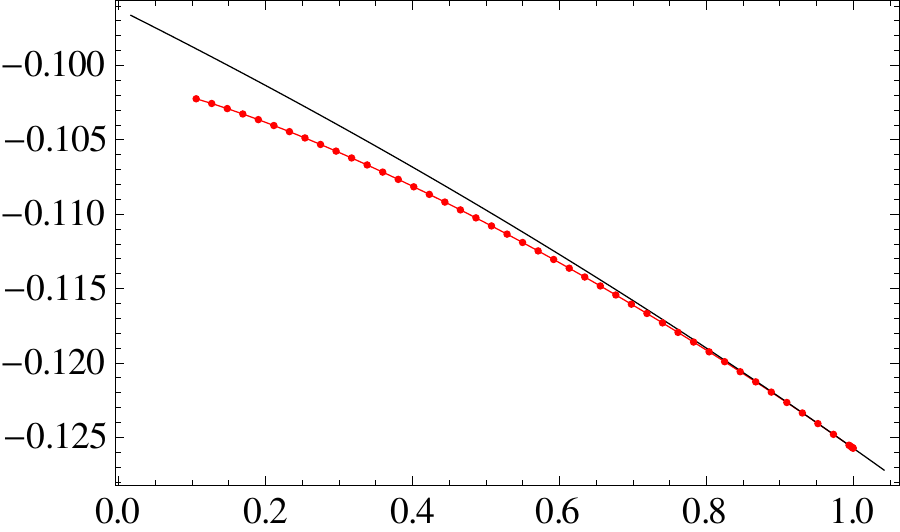}
\hspace{0.08\textwidth}
\includegraphics[width=0.4\textwidth]{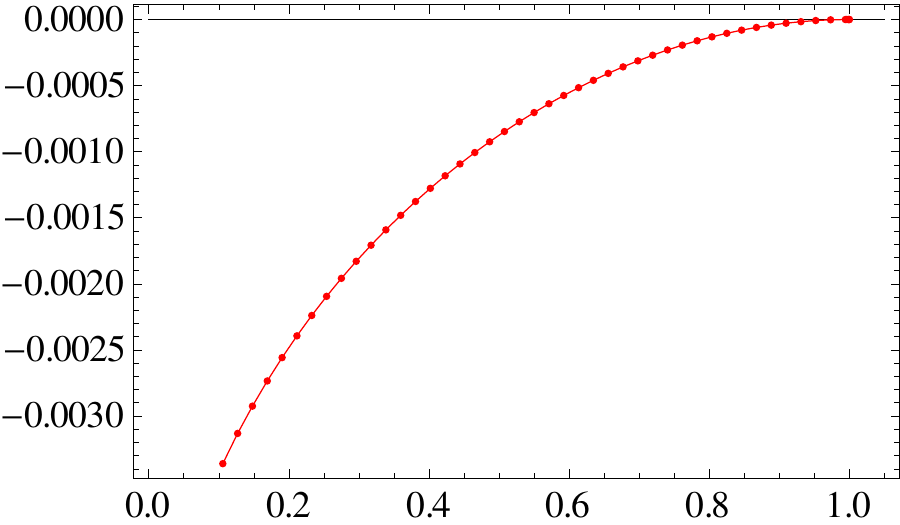}
\begin{picture}(0.1,0.1)(0,0)
\put(-400,75){\makebox(0,0){$\frac{\bar{w}}{\mu^3}$}}
\put(-300,-10){\makebox(0,0){$T/T_c$}}
\put(-80,-10){\makebox(0,0){$T/T_c$}}
\put(-195,75){\makebox(0,0){$\frac{\bar{w}-w_{RN}}{\mu^3}$}}
\end{picture}
\vskip 1em
\caption{Free energy for a fixed $k/\mu=1.1/\sqrt{2}$ branch of the model \eqref{nlmodel}, with $c_1=9.9$, indicating a continuous phase transition from the RN branch (black). \label{secondfree}}
\end{center}
\end{figure}
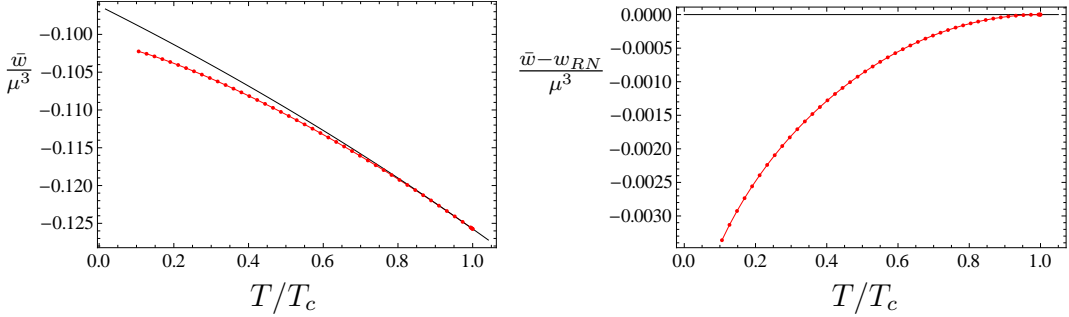

\begin{figure}[h!]
\begin{center}
\includegraphics[width=0.5\textwidth]{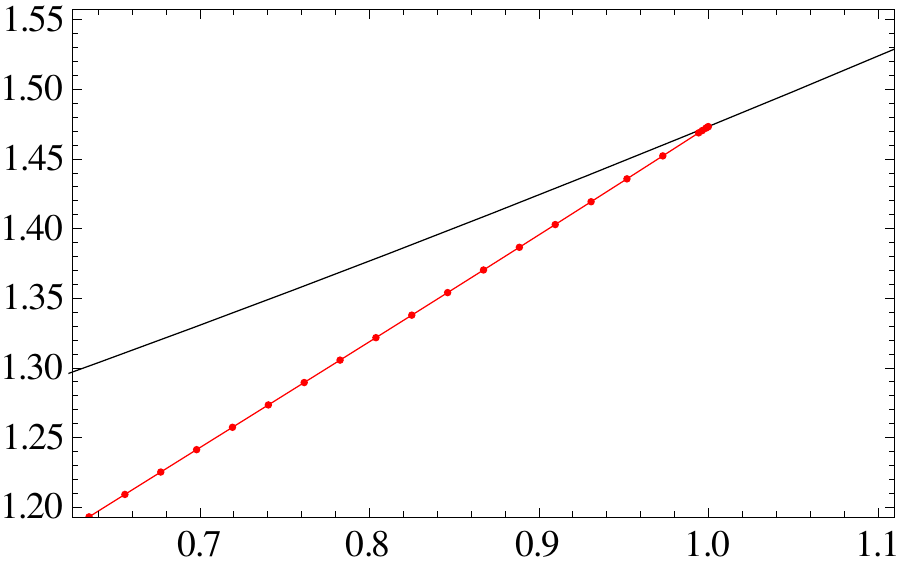}
\begin{picture}(0.1,0.1)(0,0)
\put(-110,-10){\makebox(0,0){$T/T_c$}}
\put(-240,90){\makebox(0,0){$\frac{\bar{s}}{\mu^2}$}}
\end{picture}
\vskip 1em
\caption{Entropy density for a fixed $k/\mu=1.1/\sqrt{2}$ branch of the model \eqref{nlmodel}, with $c_1=9.9$, in the vicinity of the phase transition from RN (black). Via the first law, the kink in this plot illustrates that the continuous phase transition at $(T/\mu)_c$ is second order. \label{secondentzoom}}
\end{center}
\end{figure}
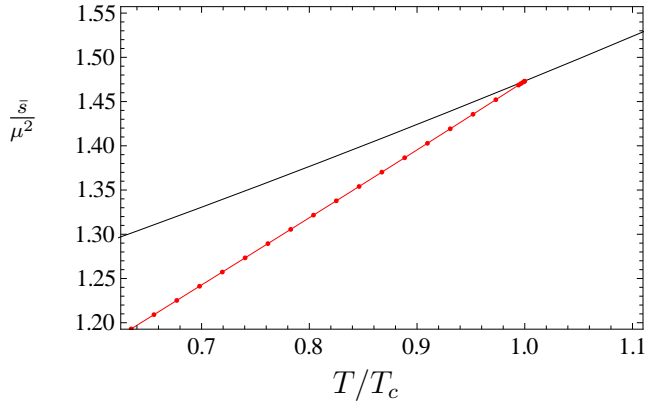

We find a second order phase transition at $T/\mu = (T/\mu)_c$. The free energy for this branch, and its difference with that of RN is shown in the panels of figure \ref{secondfree}. We find that the spatially modulated branch is thermodynamically preferred at all temperatures where it exists. To show that this corresponds to a second order phase transition, in figure \ref{secondentzoom} we plot the entropy density in the vicinity of $(T/\mu)_c$. According to the first law (which we test in section \ref{flcheck}) the entropy $\bar{s}=-\frac{\partial\bar{w}}{\partial T}$ at fixed $\mu,k$. Hence we see that the kink in the plot of $\bar{s}/\mu^2$ at the critical temperature indicates a second order phase transition.

An expression for the energy density is presented in appendix \ref{thermo}.

\subsection{Lower temperatures}
For the data presented in figure \ref{secondfree}, which we have obtained down to $T/T_c = 0.1$, we have ensured that the free energy has converged with $N$ to sufficient precision, and that the first law test is satisfied (see section \ref{flcheck}). However, the main source of error in this procedure is obtaining the energy density.\footnote{This is presumably because it appears suppressed by $z^3$ relative to the leading boundary behaviour.} If we are not interested in the free energy, then, we can go to slightly lower temperatures without compromising accuracy in any of the other observables, for example, in the entropy density $\bar{s}$. 

\begin{figure}[h!]
\begin{center}
\includegraphics[width=0.4\textwidth]{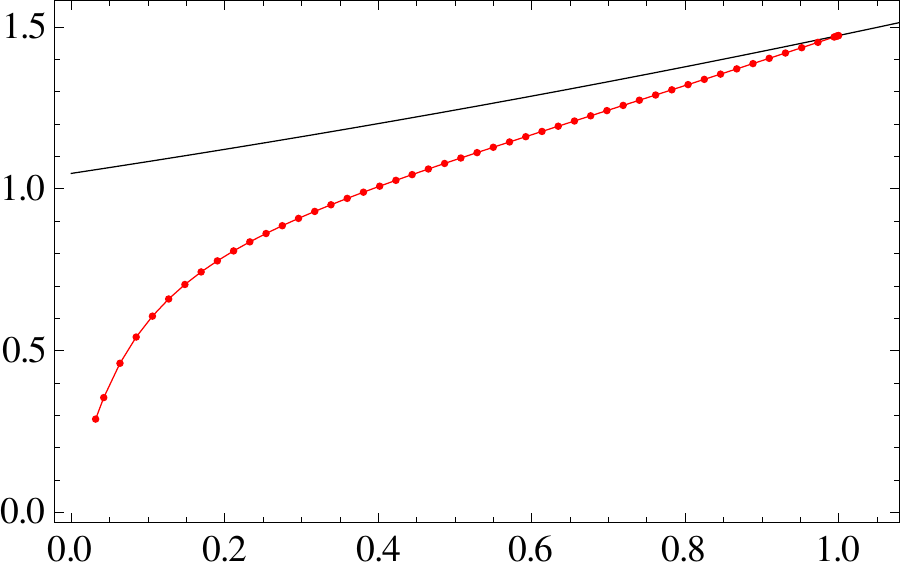}
\hspace{0.05\textwidth}
\includegraphics[width=0.4\textwidth]{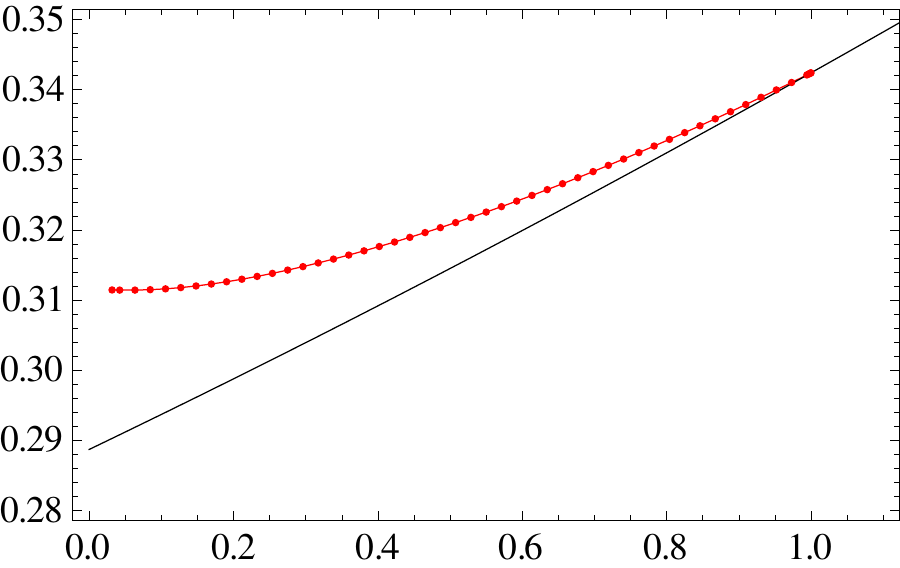}
\includegraphics[width=0.4\textwidth]{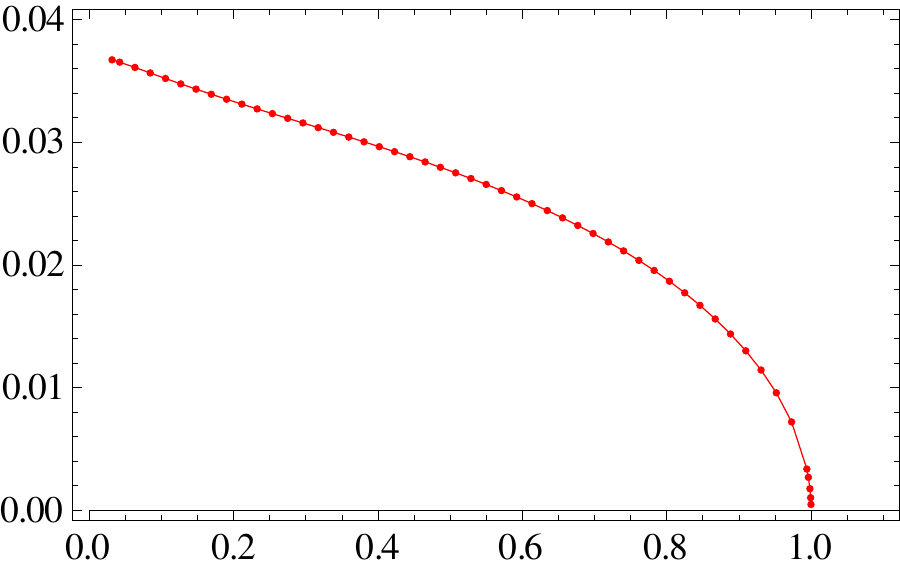}
\hspace{0.05\textwidth}
\includegraphics[width=0.4\textwidth]{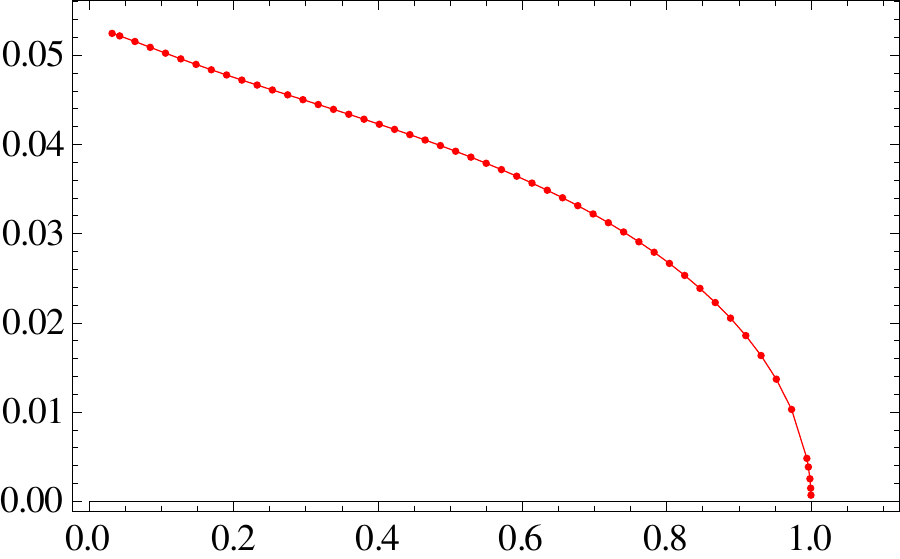}
\begin{picture}(0.1,0.1)(0,0)
\put(-85,-10){\makebox(0,0){$T/T_c$}}
\put(-290,-10){\makebox(0,0){$T/T_c$}}
\put(-390,185){\makebox(0,0){$\frac{\bar{s}}{\mu^2}$}}
\put(-185,185){\makebox(0,0){$\frac{\bar{\rho}}{\mu^2}$}}
\put(-390,70){\makebox(0,0){$\frac{v}{\mu^2}$}}
\put(-185,70){\makebox(0,0){$\frac{j}{\mu^2}$}}
\end{picture}
\vskip 1em
\caption{Various averaged quantities for a fixed $k/\mu=1.1/\sqrt{2}$ branch of the model \eqref{nlmodel}, with $c_1=9.9$, indicating an approach to an electrically charged,  spatially modulated zero-temperature state with vanishing entropy. The quantities $v$ and $j$ are defined in the text. The RN solution is in black.\label{lowt}}
\end{center}
\end{figure}
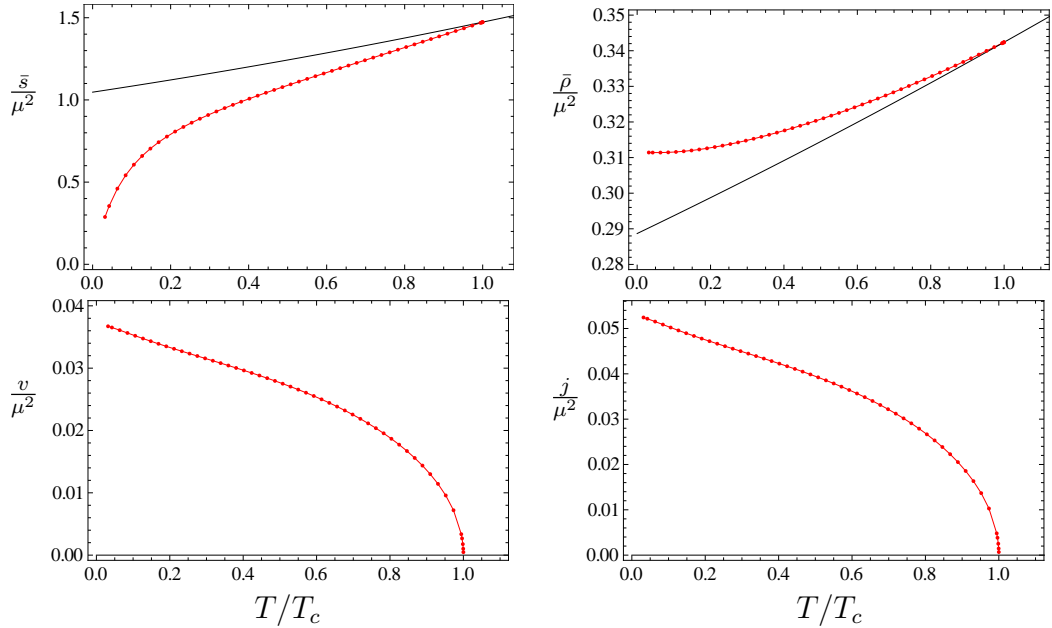

The entropy density and the charge density are presented in the upper panels of figure \ref{lowt} down to $T/T_c \simeq 0.03$. In particular we see that the system appears to approach an electrically charged,  spatially modulated infrared geometry with vanishing entropy.  The bottom two panels of figure \ref{lowt} show the quantities $v$ and $j$, defined as $v^2 \equiv \frac{k}{2\pi}\int \phi_{(2)}^2 dx$ and $j^2 \equiv \frac{k}{2\pi}\int B_{(1)}^2 dx$ where $\phi_{(2)}$ is the coefficient of $z^2$ in a boundary expansion of $\phi$ and gives the scalar VEV, whilst $B_{(1)}$ is the coefficient of $z$ in a boundary expansion of $B$ and gives the current.

\section{The model \eqref{linmodel}\label{secfirst}}
In this section we comment on a different bulk model \eqref{linmodel}, where we find that a continuous phase transition (at fixed $k/\mu$) is absent at our parameter choice.
For comparison with the results of section \ref{secondsec} we work with $c_1=9.9,n=36$ and study the fixed $k/\mu=1.1/\sqrt{2}$.
This model is obtained simply by expanding the functions $V, \tau, \vartheta$  of the model \eqref{nlmodel} to first non-trivial order in $\phi$. Consequently, a linear analysis yields identical results, e.g. for the critical temperature $(T/\mu)_c  \simeq 0.0236$. 

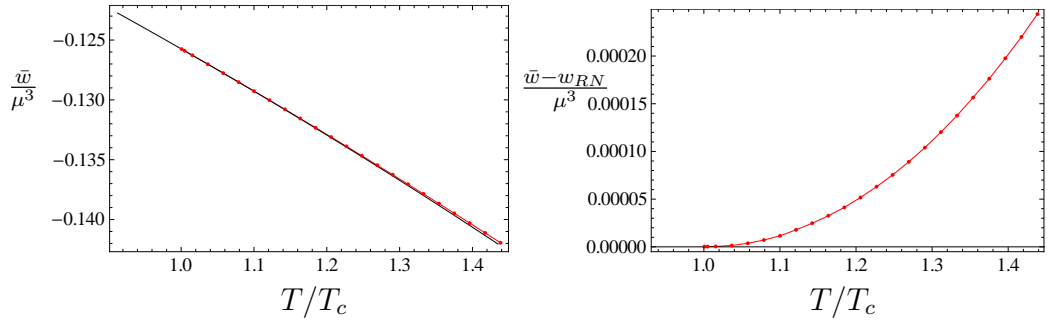
\begin{figure}[h!]
\begin{center}
\includegraphics[width=0.4\textwidth]{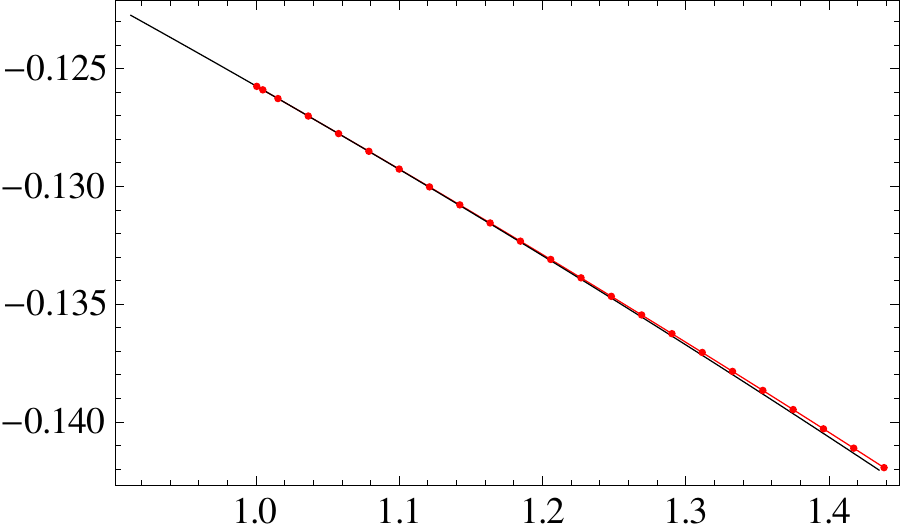}
\hspace{0.05\textwidth}
\includegraphics[width=0.4\textwidth]{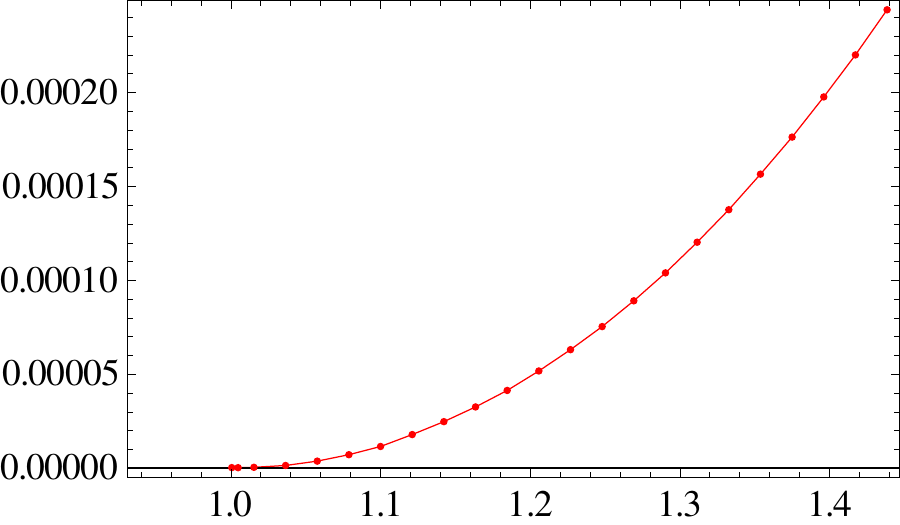}
\begin{picture}(0.1,0.1)(0,0)
\put(-390,70){\makebox(0,0){$\frac{\bar{w}}{\mu^3}$}}
\put(-280,-10){\makebox(0,0){$T/T_c$}}
\put(-80,-10){\makebox(0,0){$T/T_c$}}
\put(-185,70){\makebox(0,0){$\frac{\bar{w}-w_{RN}}{\mu^3}$}}
\end{picture}
\vskip 1em
\caption{Free energy for the fixed $k/\mu=1.1/\sqrt{2}$ branch in the model \eqref{linmodel} at $c_1=9.9, n=36$, demonstrating the absence of a continuous phase transition from the RN branch (black).\label{retrograde}}
\end{center}
\end{figure}

\begin{figure}[h!]
\begin{center}
\includegraphics[width=0.5\textwidth]{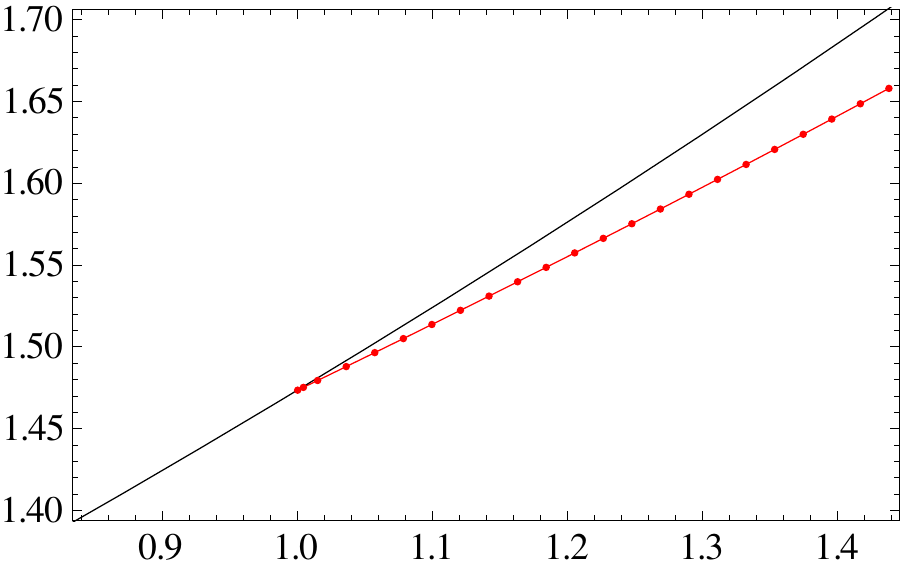}
\begin{picture}(0.1,0.1)(0,0)
\put(-110,-10){\makebox(0,0){$T/T_c$}}
\put(-240,90){\makebox(0,0){$\frac{\bar{s}}{\mu^2}$}}
\end{picture}
\vskip 1em
\caption{Entropy for the fixed $k/\mu=1.1/\sqrt{2}$ branch in the model \eqref{linmodel} at $c_1=9.9, n=36$. The RN entropy is shown in black.\label{retroentro}}
\end{center}
\end{figure}
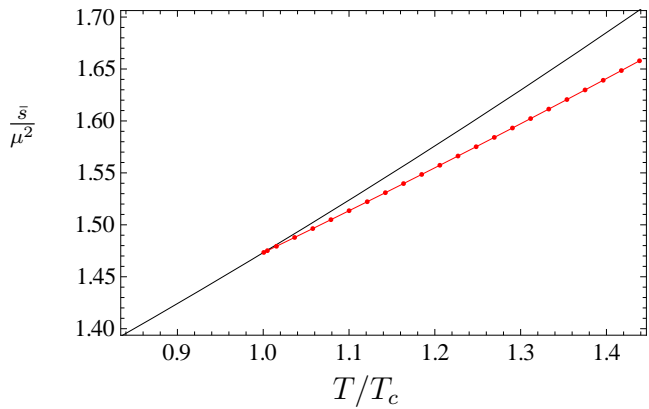

Perhaps unsurprisingly, we find drastically different behaviour to the model \eqref{nlmodel} once we move away from $(T/\mu)_c$. In particular we find that the branch of solutions emerging from the RN branch exist initially at $T/\mu>(T/\mu)_c$ and are subdominant. This is illustrated by plots of the free energy in figure \ref{retrograde}, near the phase transition. For completeness we show the entropy in figure \ref{retroentro}.

\section{Discussion\label{discussion}}
In this paper we have constructed new inhomogeneous black brane solutions in AdS, describing spatially modulated phases of a dual field theory at finite chemical potential. These modulated phases arise in the dual CFTs as new branches that spontaneously break translation invariance. We studied branches of striped black branes at fixed $k/\mu$ in two models finding a range of possible behaviours.

In the first model (section \ref{secondsec}), as the system is cooled we find a second order phase transition to the striped phase. This is our main result. Additionally, at low temperatures we have found indications that a zero entropy, electrically charged, inhomogeneous state is being approached. We note the similar behaviour of these striped solutions to the helical case \cite{Donos:2012gg,Donos:2012wi} exhibiting a zero entropy, homogeneous modulated state at $T=0$.  It would be interesting to go to lower temperatures and investigate this `ground state' more thoroughly, as well as to construct it directly at zero temperature. 

In the second model (section \ref{secfirst}), taken to be identical to the first in linear theory, we illustrated the existence of a branch of striped solutions at higher free energy than the RN branch. Moreover, these solutions exist at higher temperatures than the point at which the two branches meet, $(T/\mu)_c$. There is therefore no continuous phase transition. The natural question is whether a first order transition occurs from RN for $T/\mu>(T/\mu)_c$ \emph{e.g.} with a swallow-tail structure. We note that this possibility is distinct from the first order transitions presented in \cite{UBC}, where the transition temperature was found to be lower than $(T/\mu)_c$. If no first order transition occurs then the situation becomes reminiscent of the `retrograde' branches of holographic superconductors\cite{Buchel:2009ge,Aprile:2011uq,Donos:2011ut}. Irrespective of what happens for this particular model further along the branch, we have shown that a continuous phase transition need not occur and that, as expected, the nonlinear details of the model are important away from the critical temperature.


Finally, it would be interesting to relax the two main assumptions adopted in this work, namely,  homogeneity in the second spatial direction, $y$, and the restriction of fixed $k/\mu$. In particular, it may be that solutions with lower symmetry dominate the ensemble, and that the thermodynamically preferred periodicity changes with the temperature, as in \cite{Donos:2012gg,Donos:2012wi}.

\subsection*{Acknowledgments}
We acknowledge useful discussions with Aristomenis Donos, Jerome Gauntlett,  Simon Gentle, Blaise Gouteraux and Toby Wiseman. We thank Julian Sonner and Toby Wiseman for comments on the manuscript. Additionally we would like to thank the Perimeter Institute and Nordita for hospitality while this work was being completed. This work is supported by a Royal Commission for the Exhibition of 1851 Science Research Fellowship.
\appendix
\section{Numerical tests\label{numericaltests}}
\subsection{Testing convergence with $N$\label{nconv}}
\begin{figure}[h!]
\begin{center}
\begin{picture}(0.1,0.1)(0,0)
\put(110,-7){\makebox(0,0){$N$}}
\put(315,-7){\makebox(0,0){$N$}}
\put(265,105){\makebox(0,0){log$_{10}(|\xi^2|_{max})$}}
\put(68,101){\makebox(0,0){log$_{10} (1-\frac{\bar{\rho}_{N-1}}{\bar{\rho}_{N}})$}}
\end{picture}
\includegraphics[width=0.45\textwidth]{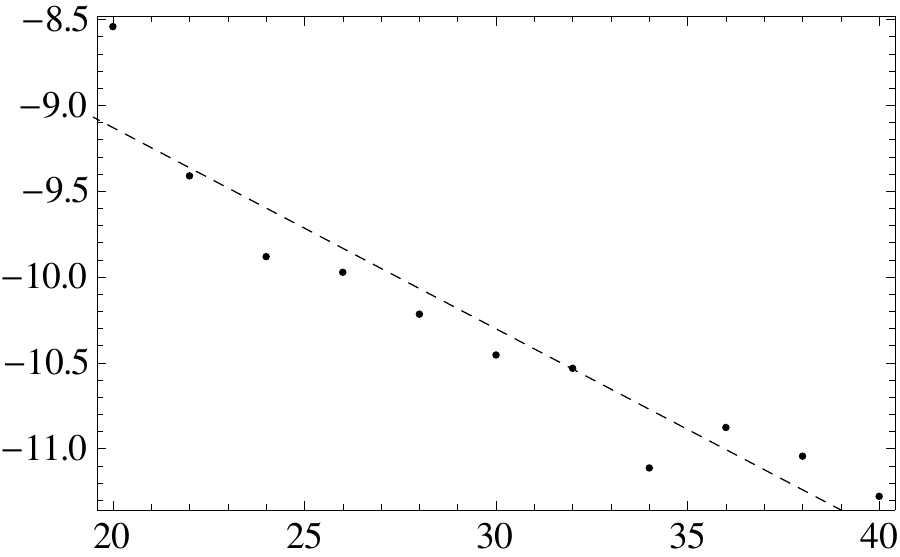}
\includegraphics[width=0.45\textwidth]{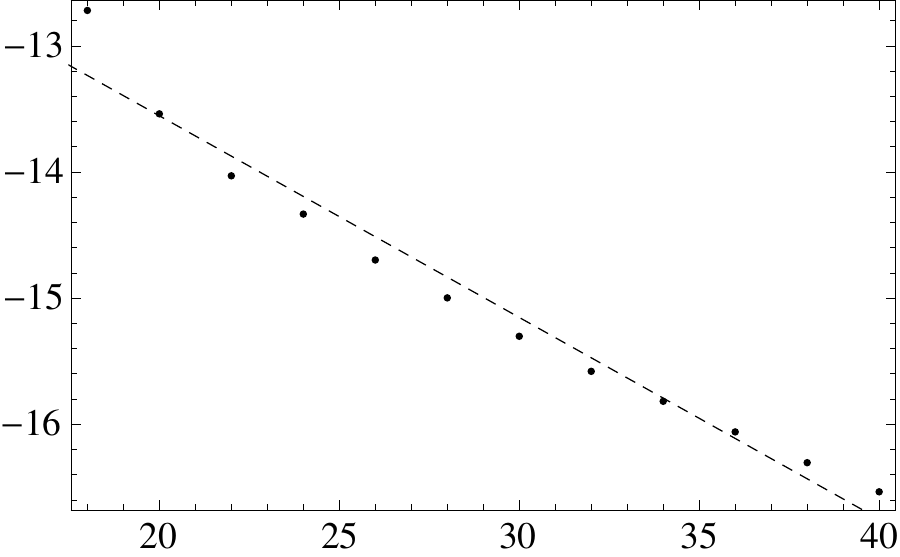}
\vskip 1em
\caption{\emph{Left:} convergence of the charge density $\bar{\rho}$ with the grid size $N$. \emph{Right:} convergence of the maximum value of $|\xi^2|$ on the grid towards zero with increasing grid size $N$. Dashed lines are the best linear fit.
  \label{nscaling}}
\end{center}
\end{figure}
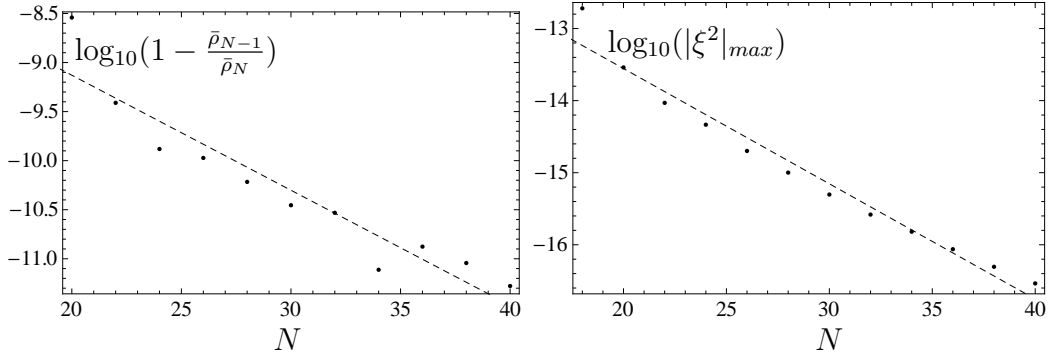

Using the solution presented in section \ref{repsec} we will illustrate the convergence properties as we scale the number of points in the grid, $N(N+1)$. As we are using spectral methods we should find an exponential rate of convergence with $N$. This is illustrated for the average of the charge density $\bar{\rho}$ in the left panel of figure \ref{nscaling}. Crucially, we must ensure that the maximum value of $|\xi^2|$ also converges appropriately towards zero so that we have a solution to the Einstein equations -- this convergence is indeed seen and is demonstrated in the right panel of figure \ref{nscaling}.

\subsection{Tests of the first law\label{flcheck}}
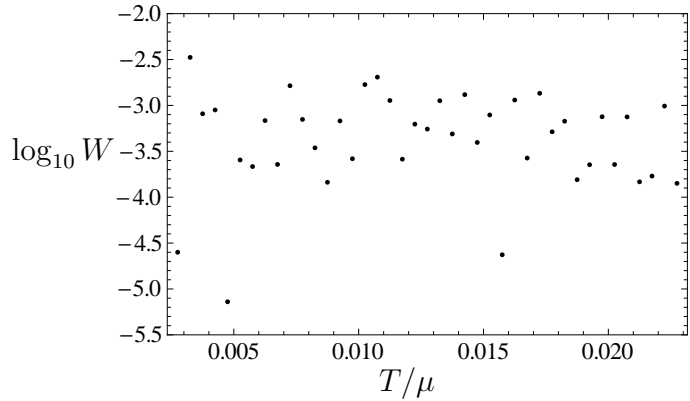
\begin{figure}[h!]
\begin{center}
\includegraphics[width=0.5\textwidth]{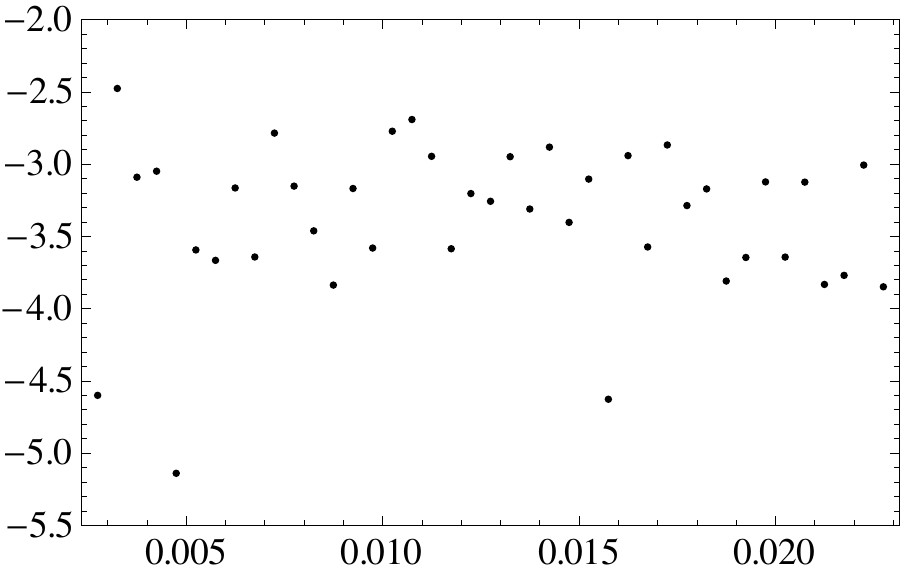}
\begin{picture}(0.1,0.25)(0,0)
\put(-240,80){\makebox(0,0){$\log_{10}W$}}
\put(-110,-7){\makebox(0,0){$T/\mu$}}
\end{picture}
\vskip 1em
\caption{Numerical check of the first law $W$ for a set of solutions on a fixed $k/\mu$ branch. Note that because $W$ is approximated here by differencing solutions with separation $\lambda = \delta\,T/\mu$, it is limited by $\lambda$ rather than $N$, and as required we find $W \sim \lambda\sim 10^{-3}$.
  \label{firstlawcheck}}
\end{center}
\end{figure}

Combining the local free energy $w(x) = \epsilon(x) - T s(x) - \mu \rho(x)$ with the first law $d \epsilon(x) =  T ds(x) + \mu d\rho(x)$ we conclude that along our fixed $k/\mu$ branch of solutions the quantity $W\equiv\frac{d \bar{w}/\mu^3}{d T/\mu}+\bar{s}/\mu^2$ must vanish. In particular, approximating the derivatives by the difference of two nearby solutions with parametric separation $\lambda = \delta\,T/\mu$, we should expect to find numerically that $W = 0+ O(\lambda)$. Indeed, in figure \ref{firstlawcheck} we see that this condition is met along the set of solutions of section \ref{secondsec}, where $\lambda \sim 10^{-3}$.

\section{Free energy \label{thermo}}
To calculate the free energy we follow \cite{Balasubramanian:1999re,Donos:2012yu}. Including boundary terms, the action for this system is given by,
\be
S = S_{\text{bulk}}  - \int d^3x \sqrt{-h} (K-4+\phi^2),
\ee
where $S_{\text{bulk}}$ is given by \eqref{sbulk}, 
$h_{ab}$ is the induced metric on the boundary 
and $K_{ab}$ the extrinsic curvature. 
From this we may extract the renormalised boundary stress tensor $T_{ab} = \frac{2}{\sqrt{-h}}\frac{\delta S}{\delta h^{ab}}$, which is given in terms of the data appearing in the boundary expansion. From $T_{tt}$ we may identify the local energy density,
\be
\epsilon(x) = 2+\frac{\mu^2}{2} - 3 T^{(3)}(x)
\ee
where $T^{(3)}(x) = \frac{1}{3!}\partial_z^3 T\big|_{z=0}$.
This allows us to evaluate a local expression for the thermodynamic potential $w(x) = \epsilon(x) - T s(x) - \mu \rho(x)$. The entropy density is given in the usual way by $s(x) =\frac{a(x)}{4G}$ where $a(x)$ is the area measure of the horizon. For the thermodynamical properties of the solutions presented in this paper we work with the averaged quantities, $\bar{w}, \bar{\epsilon}, \bar{s}, \bar{\rho}$, defined by \eqref{average}.

\bibliographystyle{utphys}
\bibliography{stripes}{}

\providecommand{\href}[2]{#2}\begingroup\raggedright\begin{thebibliography}{10}

\bibitem{Gubser:2008px}
S.~S. Gubser, ``{Breaking an Abelian Gauge Symmetry Near a Black Hole
  Horizon},'' \href{http://dx.doi.org/10.1103/PhysRevD.78.065034}{{\em Phys.
  Rev.} {\bfseries D78} (2008) 065034},
\href{http://arxiv.org/abs/0801.2977}{{\ttfamily arXiv:0801.2977 [hep-th]}}.

\bibitem{Hartnoll:2008vx}
S.~A. Hartnoll, C.~P. Herzog, and G.~T. Horowitz, ``{Building a Holographic
  Superconductor},''
  \href{http://dx.doi.org/10.1103/PhysRevLett.101.031601}{{\em Phys. Rev.
  Lett.} {\bfseries 101} (2008) 031601},
\href{http://arxiv.org/abs/0803.3295}{{\ttfamily arXiv:0803.3295 [hep-th]}}.

\bibitem{Hartnoll:2008kx}
S.~A. Hartnoll, C.~P. Herzog, and G.~T. Horowitz, ``{Holographic
  Superconductors},''
  \href{http://dx.doi.org/10.1088/1126-6708/2008/12/015}{{\em JHEP} {\bfseries
  12} (2008) 015},
\href{http://arxiv.org/abs/0810.1563}{{\ttfamily arXiv:0810.1563 [hep-th]}}.

\bibitem{Donos:2011ut}
A.~Donos and J.~P. Gauntlett, ``{Superfluid black branes in $AdS_4\times
  S^7$},'' \href{http://dx.doi.org/10.1007/JHEP06(2011)053}{{\em JHEP}
  {\bfseries 06} (2011) 053},
\href{http://arxiv.org/abs/1104.4478}{{\ttfamily arXiv:1104.4478 [hep-th]}}.

\bibitem{Gauntlett:2009dn}
J.~P. Gauntlett, J.~Sonner, and T.~Wiseman, ``{Holographic superconductivity in
  M-Theory},'' \href{http://dx.doi.org/10.1103/PhysRevLett.103.151601}{{\em
  Phys. Rev. Lett.} {\bfseries 103} (2009) 151601},
\href{http://arxiv.org/abs/0907.3796}{{\ttfamily arXiv:0907.3796 [hep-th]}}.

\bibitem{Gubser:2009qm}
S.~S. Gubser, C.~P. Herzog, S.~S. Pufu, and T.~Tesileanu, ``{Superconductors
  from Superstrings},''
  \href{http://dx.doi.org/10.1103/PhysRevLett.103.141601}{{\em Phys. Rev.
  Lett.} {\bfseries 103} (2009) 141601},
\href{http://arxiv.org/abs/0907.3510}{{\ttfamily arXiv:0907.3510 [hep-th]}}.

\bibitem{Nakamura:2009tf}
S.~Nakamura, H.~Ooguri, and C.-S. Park, ``{Gravity Dual of Spatially Modulated
  Phase},'' \href{http://dx.doi.org/10.1103/PhysRevD.81.044018}{{\em Phys.
  Rev.} {\bfseries D81} (2010) 044018},
\href{http://arxiv.org/abs/0911.0679}{{\ttfamily arXiv:0911.0679 [hep-th]}}.

\bibitem{Ooguri:2010kt}
H.~Ooguri and C.-S. Park, ``{Holographic End-Point of Spatially Modulated Phase
  Transition},'' \href{http://dx.doi.org/10.1103/PhysRevD.82.126001}{{\em Phys.
  Rev.} {\bfseries D82} (2010) 126001},
\href{http://arxiv.org/abs/1007.3737}{{\ttfamily arXiv:1007.3737 [hep-th]}}.

\bibitem{Donos:2011bh}
A.~Donos and J.~P. Gauntlett, ``{Holographic striped phases},''
  \href{http://dx.doi.org/10.1007/JHEP08(2011)140}{{\em JHEP} {\bfseries 1108}
  (2011) 140},
\href{http://arxiv.org/abs/1106.2004}{{\ttfamily arXiv:1106.2004 [hep-th]}}.

\bibitem{Donos:2011ff}
A.~Donos and J.~P. Gauntlett, ``{Holographic helical superconductors},''
  \href{http://dx.doi.org/10.1007/JHEP12(2011)091}{{\em JHEP} {\bfseries 12}
  (2011) 091},
\href{http://arxiv.org/abs/1109.3866}{{\ttfamily arXiv:1109.3866 [hep-th]}}.

\bibitem{Donos:2011qt}
A.~Donos, J.~P. Gauntlett, and C.~Pantelidou, ``{Spatially modulated
  instabilities of magnetic black branes},''
  \href{http://dx.doi.org/10.1007/JHEP01(2012)061}{{\em JHEP} {\bfseries 01}
  (2012) 061},
\href{http://arxiv.org/abs/1109.0471}{{\ttfamily arXiv:1109.0471 [hep-th]}}.

\bibitem{Donos:2011pn}
A.~Donos, J.~P. Gauntlett, and C.~Pantelidou, ``{Magnetic and Electric AdS
  Solutions in String- and M-Theory},''
  \href{http://dx.doi.org/10.1088/0264-9381/29/19/194006}{{\em
  Class.Quant.Grav.} {\bfseries 29} (2012) 194006},
\href{http://arxiv.org/abs/1112.4195}{{\ttfamily arXiv:1112.4195 [hep-th]}}.

\bibitem{Vegh:2013sk}
D.~Vegh, ``{Holography without translational symmetry},''
\href{http://arxiv.org/abs/1301.0537}{{\ttfamily arXiv:1301.0537 [hep-th]}}.

\bibitem{Donos:2013gda}
A.~Donos and J.~P. Gauntlett, ``{Holographic charge density waves},''
\href{http://arxiv.org/abs/1303.4398}{{\ttfamily arXiv:1303.4398 [hep-th]}}.

\bibitem{Domokos:2007kt}
S.~K. Domokos and J.~A. Harvey, ``{Baryon number-induced Chern-Simons couplings
  of vector and axial-vector mesons in holographic QCD},''
  \href{http://dx.doi.org/10.1103/PhysRevLett.99.141602}{{\em Phys. Rev. Lett.}
  {\bfseries 99} (2007) 141602},
\href{http://arxiv.org/abs/0704.1604}{{\ttfamily arXiv:0704.1604 [hep-ph]}}.

\bibitem{Ooguri:2010xs}
H.~Ooguri and C.-S. Park, ``{Spatially Modulated Phase in Holographic
  Quark-Gluon Plasma},''
  \href{http://dx.doi.org/10.1103/PhysRevLett.106.061601}{{\em Phys. Rev.
  Lett.} {\bfseries 106} (2011) 061601},
\href{http://arxiv.org/abs/1011.4144}{{\ttfamily arXiv:1011.4144 [hep-th]}}.

\bibitem{Bayona:2011ab}
C.~A.~B. Bayona, K.~Peeters, and M.~Zamaklar, ``{A non-homogeneous ground state
  of the low-temperature Sakai-Sugimoto model},''
  \href{http://dx.doi.org/10.1007/JHEP06(2011)092}{{\em JHEP} {\bfseries 06}
  (2011) 092},
\href{http://arxiv.org/abs/1104.2291}{{\ttfamily arXiv:1104.2291 [hep-th]}}.

\bibitem{Bergman:2011rf}
O.~Bergman, N.~Jokela, G.~Lifschytz, and M.~Lippert, ``{Striped instability of
  a holographic Fermi-like liquid},''
  \href{http://dx.doi.org/10.1007/JHEP10(2011)034}{{\em JHEP} {\bfseries 10}
  (2011) 034},
\href{http://arxiv.org/abs/1106.3883}{{\ttfamily arXiv:1106.3883 [hep-th]}}.

\bibitem{Iizuka:2012iv}
N.~Iizuka, S.~Kachru, N.~Kundu, P.~Narayan, N.~Sircar, {\em et al.}, ``{Bianchi
  Attractors: A Classification of Extremal Black Brane Geometries},''
\href{http://arxiv.org/abs/1201.4861}{{\ttfamily arXiv:1201.4861 [hep-th]}}.

\bibitem{Donos:2012gg}
A.~Donos and J.~P. Gauntlett, ``{Helical superconducting black holes},''
  \href{http://dx.doi.org/10.1103/PhysRevLett.108.211601}{{\em Phys.Rev.Lett.}
  {\bfseries 108} (2012) 211601},
\href{http://arxiv.org/abs/1203.0533}{{\ttfamily arXiv:1203.0533 [hep-th]}}.

\bibitem{Donos:2012wi}
A.~Donos and J.~P. Gauntlett, ``{Black holes dual to helical current phases},''
\href{http://arxiv.org/abs/1204.1734}{{\ttfamily arXiv:1204.1734 [hep-th]}}.

\bibitem{Headrick:2009pv}
M.~Headrick, S.~Kitchen, and T.~Wiseman, ``{A New approach to static numerical
  relativity, and its application to Kaluza-Klein black holes},''
  \href{http://dx.doi.org/10.1088/0264-9381/27/3/035002}{{\em
  Class.Quant.Grav.} {\bfseries 27} (2010) 035002},
\href{http://arxiv.org/abs/0905.1822}{{\ttfamily arXiv:0905.1822 [gr-qc]}}.

\bibitem{Adam:2011dn}
A.~Adam, S.~Kitchen, and T.~Wiseman, ``{A numerical approach to finding general
  stationary vacuum black holes},''
  \href{http://dx.doi.org/10.1088/0264-9381/29/16/165002}{{\em
  Class.Quant.Grav.} {\bfseries 29} (2012) 165002},
\href{http://arxiv.org/abs/1105.6347}{{\ttfamily arXiv:1105.6347 [gr-qc]}}.

\bibitem{Wiseman:2011by}
T.~Wiseman, ``{Numerical construction of static and stationary black holes},''
\href{http://arxiv.org/abs/1107.5513}{{\ttfamily arXiv:1107.5513 [gr-qc]}}.

\bibitem{Horowitz:2012ky}
G.~T. Horowitz, J.~E. Santos, and D.~Tong, ``{Optical Conductivity with
  Holographic Lattices},''
  \href{http://dx.doi.org/10.1007/JHEP07(2012)168}{{\em JHEP} {\bfseries 1207}
  (2012) 168},
\href{http://arxiv.org/abs/1204.0519}{{\ttfamily arXiv:1204.0519 [hep-th]}}.

\bibitem{Horowitz:2012gs}
G.~T. Horowitz, J.~E. Santos, and D.~Tong, ``{Further Evidence for
  Lattice-Induced Scaling},''
  \href{http://dx.doi.org/10.1007/JHEP11(2012)102}{{\em JHEP} {\bfseries 1211}
  (2012) 102},
\href{http://arxiv.org/abs/1209.1098}{{\ttfamily arXiv:1209.1098 [hep-th]}}.

\bibitem{Horowitz:2013jaa}
G.~T. Horowitz and J.~E. Santos, ``{General Relativity and the Cuprates},''
\href{http://arxiv.org/abs/1302.6586}{{\ttfamily arXiv:1302.6586 [hep-th]}}.

\bibitem{UBC}
M.~Rozali, D.~Smyth, E.~Sorkin, and J.~B. Stang, ``{Holographic Stripes},''
\href{http://arxiv.org/abs/1211.5600}{{\ttfamily arXiv:1211.5600 [hep-th]}}.

\bibitem{DonosSolo}
A.~Donos, ``{Striped phases from holography},''
\href{http://arxiv.org/abs/1303.7211}{{\ttfamily arXiv:1303.7211 [hep-th]}}.

\bibitem{Gauntlett:2009bh}
J.~P. Gauntlett, J.~Sonner, and T.~Wiseman, ``{Quantum Criticality and
  Holographic Superconductors in M- theory},''
  \href{http://dx.doi.org/10.1007/JHEP02(2010)060}{{\em JHEP} {\bfseries 02}
  (2010) 060},
\href{http://arxiv.org/abs/0912.0512}{{\ttfamily arXiv:0912.0512 [hep-th]}}.

\bibitem{Gauntlett:2009zw}
J.~P. Gauntlett, S.~Kim, O.~Varela, and D.~Waldram, ``{Consistent
  Supersymmetric Kaluza--Klein Truncations with Massive Modes},''
  \href{http://dx.doi.org/10.1088/1126-6708/2009/04/102}{{\em JHEP} {\bfseries
  04} (2009) 102},
\href{http://arxiv.org/abs/0901.0676}{{\ttfamily arXiv:0901.0676 [hep-th]}}.

\bibitem{Donos:2012yu}
A.~Donos, J.~P. Gauntlett, J.~Sonner, and B.~Withers, ``{Competing orders in
  M-theory: superfluids, stripes and metamagnetism},''
  \href{http://dx.doi.org/10.1007/JHEP03(2013)108}{{\em JHEP} {\bfseries 1303}
  (2013) 108},
\href{http://arxiv.org/abs/1212.0871}{{\ttfamily arXiv:1212.0871 [hep-th]}}.

\bibitem{Tref}
L.~N. Trefethen, {\em Spectral methods in MatLab}.
\newblock SIAM, Philadelphia, PA, USA, 2000.

\bibitem{Buchel:2009ge}
A.~Buchel and C.~Pagnutti, ``{Exotic Hairy Black Holes},''
  \href{http://dx.doi.org/10.1016/j.nuclphysb.2009.08.017}{{\em Nucl.Phys.}
  {\bfseries B824} (2010) 85--94},
\href{http://arxiv.org/abs/0904.1716}{{\ttfamily arXiv:0904.1716 [hep-th]}}.

\bibitem{Aprile:2011uq}
F.~Aprile, D.~Roest, and J.~G. Russo, ``{Holographic Superconductors from
  Gauged Supergravity},'' \href{http://dx.doi.org/10.1007/JHEP06(2011)040}{{\em
  JHEP} {\bfseries 1106} (2011) 040},
\href{http://arxiv.org/abs/1104.4473}{{\ttfamily arXiv:1104.4473 [hep-th]}}.

\bibitem{Balasubramanian:1999re}
V.~Balasubramanian and P.~Kraus, ``{A stress tensor for anti-de Sitter
  gravity},'' \href{http://dx.doi.org/10.1007/s002200050764}{{\em Commun. Math.
  Phys.} {\bfseries 208} (1999) 413--428},
\href{http://arxiv.org/abs/hep-th/9902121}{{\ttfamily arXiv:hep-th/9902121}}.

\end{thebibliography}\endgroup

\end{document}